\documentclass[%
twocolumn,
 amsmath,amssymb,
 aps,
]{revtex4-2}

\usepackage{graphicx}
\usepackage{dcolumn}
\usepackage{bm}
\usepackage{hyperref}
\usepackage[capitalise,nameinlink]{cleveref}
\usepackage{ulem}
\usepackage{multirow}
\usepackage{array}
\usepackage{xr}

\newcolumntype{P}{>{\centering\arraybackslash}m{2cm}}

\usepackage{xcolor}
\usepackage{slashed}
\definecolor{myred}{RGB}{255,0,0}
\definecolor{myblue}{RGB}{0,0,255}
\definecolor{mypurple}{RGB}{153,102,204}

\newcommand{\dla}[1] {$d$-$\Lambda${\bf #1}}

\newcommand{\pla}[1] {$p$-$\Lambda${\bf #1}}

\begin{document}

\preprint{}

\title{First observation of deuteron-{$\Lambda$} correlations at RHIC}

\author{}

\collaboration{STAR Collaboration}

\date{\today}

\begin{abstract}
Precise experimental information on hyperon-nucleon interactions is scarce but of paramount importance to our understanding of the inner structure of compact stars. In this letter, 
we report the first experimental results of correlation functions between deuterons ($d$) and $\Lambda$ hyperons in Au+Au collisions at $\sqrt{s_{NN}}$ = 3.0\,GeV measured by the STAR experiment at RHIC. A clear enhancement at small relative momenta has been observed in the correlation function. 
Through a Bayesian inference analysis, the source size parameters as a function of collision centrality and the spin-dependent strong interaction parameters (scattering length $f_0$ and effective range $d_0$) are extracted using the Lednick\'y-Lyuboshitz formalism. The derived doublet spin state parameters 
($f_0,d_0$) lead to a novel method to precisely determine $\Lambda$ separation energy for the weakly bounded hypertriton $_{\Lambda}^{3}{\mathrm{H}}$.
\end{abstract}

\maketitle

Nucleon-nucleon ($N$-$N$) and hyperon-nucleon ($Y$-$N$) interactions are key ingredients to understand \mbox{(hyper-)nuclei} structure and equation-of-state (EoS) of nuclear matter under an extreme dense environment, thus may offer insights into the inner structure of compact stars~\cite{Lattimer:2004pg, Tolos:2020aln, Le:2024rkd}.
Historically, interactions between nucleons have been studied using low energy nucleon scattering experiments along with the properties of nuclear bound states, and analyzed through effective range expansion (ERE) theory~\cite{Bethe:1949yr, Kharchenko:1990qbn, Epelbaum:2008ga}, in which strong interactions can be characterized by scattering length and effective range~\cite{Hackenburg:2006qd}. While there have been sufficient data to constrain these parameters well for $N$-$N$ interactions, this is not the case for $Y$-$N$ interactions~\cite{Hackenburg:2006qd,Alexander:1968acu}.

In recent years, strong interactions between different baryons and mesons have been investigated using the femtoscopic correlation technique in heavy-ion and proton-proton collisions at the Relativistic Heavy Ion Collider (RHIC)~\cite{STAR:2015kha, STAR:2014dcy} and the Large Hadron Collider (LHC)~\cite{Fabbietti:2020bfg}. 
The momentum space correlation functions of non-identical particles are factorized into an emission source convoluted with the interactions between particles that are produced from the source. Furthermore, this strong interaction can be characterized by the scattering length and effective range parameters using the ERE framework~\cite{Lednicky:1981su, Fabbietti:2020bfg}.
Such a technique has been successfully applied to the colliding systems and the extracted strong interaction parameters from proton-proton and antiproton-antiproton pair correlations are consistent with those extracted from the low energy scattering data~\cite{STAR:2015kha, Wiringa:1994wb,Xu:2024dnd}.

Correlations with hyperons (e.g. $\Lambda$) are expected to offer direct information on the $Y$-$N$ strong interactions. Measurement of the proton-$\Lambda$ (\pla{}) correlations from STAR~\cite{STAR:2005rpl} and ALICE~\cite{ALICE:2018ysd} yielded a much smaller scattering length compared to proton-proton scattering length~\cite{STAR:2015kha}, indicating a weaker interaction strength between hyperons and nucleons. On the other hand, correlations between protons and light nuclei in heavy-ion collisions from STAR suggested repulsive interactions between these particles and the extracted strong interaction parameters are consistent with those constrained by bound nuclei states~\cite{STAR:2024zvj}. \dla{} correlation is of particular interest since the strong scattering parameters between \dla{} can offer insights into the inner structure of hypertriton~\cite{Haidenbauer:2020uew}, and they have never been measured before. Furthermore, correlations involving deuterons (bound states of protons and neutrons) may provide inputs to the understanding of possible three-body interactions between nucleons and hyperons, which was suggested to play a critical role to the EoS of neutron stars~\cite{Gerstung:2020ktv}.

In this Letter, we report the first observation of the \dla{} correlation in Au+Au collision at center-of-mass energy $\sqrt{s_{{NN}}}=3$\,GeV with the STAR experiment at RHIC. 
The correlation function between $d$ and $\Lambda$ is defined as follows,
\begin{equation}
    C(\mathbf{p_{d}}, \mathbf{p_{\Lambda}}) \equiv \frac{P(\mathbf{p_{d}},\mathbf{p_{\Lambda}})}{P(\mathbf{p_{d}})\cdot P(\mathbf{p_{\Lambda}})},
    \label{cf_statistical}
\end{equation}
where $\mathbf{p_{d}}$ and $\mathbf{p_{\Lambda}}$ are the momentum of $d$ and $\Lambda$, respectively. $P$ is the probability that such particles (or pairs) are produced at a given momentum~\cite{Lisa:2005dd}. 
In the experimental measurement, the following equation is used to obtain the correlation function,
\begin{equation}
    C({k^*}) = \mathcal{N}\frac{A({k^*})}{B({k^*})}. 
    \label{cf_experimental}
\end{equation}
Here, ${k^*}$ is the particle momentum difference in the pair rest frame. $A({k^*})$ is the distribution of ${k^*}$ with both particles from the same event, $B({k^*})$ is for two particles from different events by mixed-event technique, and $\mathcal{N}$ is the normalization factor~\cite{Lisa:2005dd}. $C({k^*})$ at unity indicates non-correlation scenario.

The dataset used in the analysis was collected by STAR under the fixed-target (FXT) setup with a gold beam of energy 3.85\,GeV/nucleon bombarding a 0.25\,mm thin gold target. The target is located $\sim$210\,cm to the west of the center of the STAR detector~\cite{STAR:2021ozh, STAR:2021hyx}.
The main detectors used in this study are the Time Projection Chamber (TPC)~\cite{Anderson:2003ur} and the Time of Flight (TOF) detector~\cite{ Llope:2012zz, Chen:2024aom}, which cover the pseudorapidity ($\eta$) range of $0.1<\eta<2$ and $0.1<\eta<1.5$, respectively, with full azimuthal coverage for both.
After requiring the collision vertex position of each event to be within $\pm 2$\,cm of the known target position along the beam direction ($V_z$), and within a radius of 1.5\,cm perpendicular to the beam direction, around 250 million events are used for further analysis. 
In this analysis, we required at least 23 hits (out of 45) to reconstruct a track and identify as a particle in the TPC. 
The centrality is determined using the charged particle multiplicity distribution within the TPC acceptance matched to Monte Carlo (MC) Glauber calculations~\cite{STAR:2021ozh, STAR:2021fge, Ray:2007av}.

\begin{figure}[ht]
    \centering    
    \includegraphics[width=0.45\textwidth]{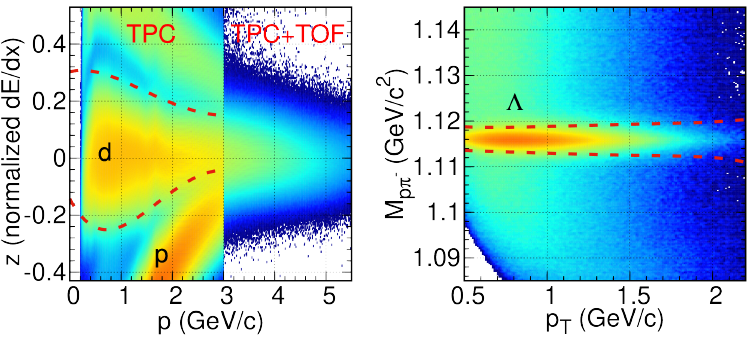}
    \caption{(Left) Selection of deuterons by TPC $z$ (vertical axis) at different momentum ($p$). TPC is used when $p<3$\,GeV/$c$, and TOF information additionally constrains the selection to remove protons when $p>3$\,GeV/$c$. (Right) The $p-\pi^-$  invariant mass distribution for the reconstructed $\Lambda$ candidates vs. $p_{T}$. Red curves show the selection cuts. }
    \label{fig:dL_acceptance}
\end{figure}

The ionization energy loss of the particles in TPC~\cite{Bichsel:2006cs} is used to identify deuterons below 3\,GeV/$c$. 
 TOF information is used to aid deuteron identification above 3 GeV/$c$.  The normalized energy loss variable $z\equiv \ln \left[ \langle dE/dx \rangle/\langle dE/dx \rangle_{\rm Bichsel}\right]$~\cite{STAR:2021ozh} in the TPC gas is used to separate deuterons from protons and tritons at different momentum as shown in Fig.~\ref{fig:dL_acceptance} left panel. Greater than 96\% purity for deuterons is reached in all kinematic regions.
$\Lambda$ candidates are reconstructed with the KF-Particle package~\cite{Kisel:2018nvd, STAR:2020xbm} using the decay channel $\Lambda\rightarrow p\pi^{-}$ (branching ratio is 64\%)~\cite{STAR:2021ozh}. We require the selected invariant mass to be within $2\sigma$ of the peak. Figure~\ref{fig:dL_acceptance} right panel shows the $p\pi^-$ invariant mass distribution vs. transverse momentum ($p_T$). The average $\Lambda$ purity is 84\%.
Kinematic windows, rapidity $y\in [-1,0]$ and $p_T\in[0.6,3]$\,GeV/$c$ for deuterons and $y\in [-1,0]$ and $p_T\in[0.4,2.2]$\,GeV/$c$ for $\Lambda$, are used for the correlation measurement. 

\begin{figure*}[ht]
    \centering    
    \includegraphics[width=0.98\textwidth]{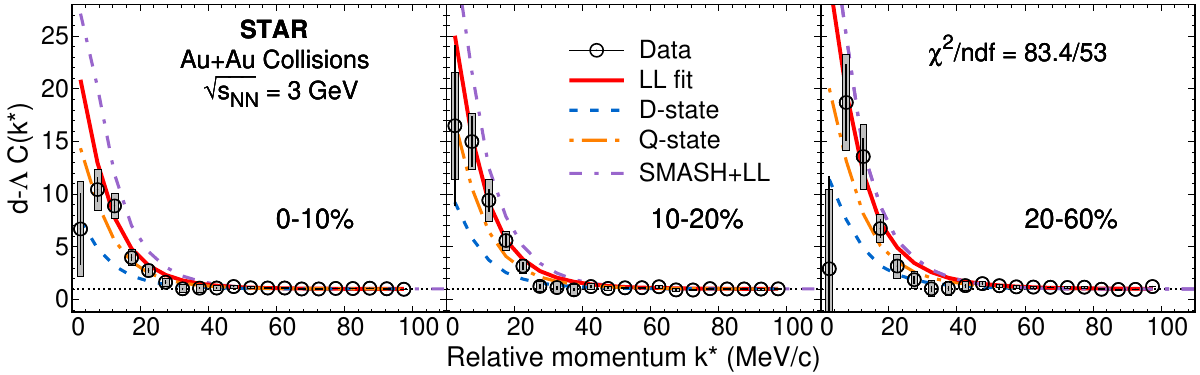}
    \caption{\dla{} correlation function ($C(k^*)$) shown as a function of relative momentum ($k^*$) at different centralities. Statistical and systematical uncertainties are shown with bars and boxes, respectively. Solid red lines are curves from the constrained fits with the LL approach (from 0 to 100 MeV/c). The combined $\chi^2/ndf$ for the three centralities is 83.4/53. Blue and orange dashed lines show the contribution from the D-state and Q-state, respectively. The dashed purple line shows the SMASH+LL model~\cite{SMASH:2016zqf,Steinheimer:2012tb}. See text for details.}
    \label{fig:dL_CF}
\end{figure*}

The \dla{} pairs counted from the same-event and mixed-event ensembles are used to calculate the correlation function via Eq.~\ref{cf_experimental}, and the region of \mbox{300 $<$ $k^{*}$ $<$ 500 MeV/$c$} is used for normalization to unity. 
A selection which requires the particle pairs have a gap in pseudorapidity ($|\Delta\eta| > 0.03$) or local azimuth ($|\Delta\phi| > 0.05$) in the TPC is applied to suppress the track merging effect from the correlation functions. The track splitting effect is found to be negligible.
The raw correlation function is corrected for the purity of the particles.
The observed \dla{} correlation contains a three-body decay contribution from hypertriton $^3_{\Lambda}{\mathrm{H}} \rightarrow {p}\pi^{-}{d}$. The mass distributions of $p\pi^-$ pairs from hypertriton three-body decay and real $\Lambda$ decay are experimentally indistinguishable due to the hypertriton decay kinematics and limited experimental mass resolution. 
This contribution is estimated using the STAR measured $^3_{\Lambda}{\rm H}$ yield~\cite{STAR:2022fnj, Chen:2023mel} folded in with realistic decay kinematics~\cite{Chen:2023mel,Kamada:1997rv}, 
which leads to 6-21\% of reconstructed \dla{} entries at $k^*< 100$ MeV/$c$ in different centralities.
Such contamination is subtracted from the \dla{} correlation measurement.
Lastly, the correlations are corrected for the feed-down contribution to $\Lambda$ from $\Sigma$ and $\Xi$ decays, which is estimated to be $(30\pm4)$\% using thermal model and transport UrQMD model results together with detector simulations~\cite{Vovchenko:2015idt, Vovchenko:2019pjl, Andronic:2017pug}.

The presented correlation function data include systematic uncertainties due to experimental and analysis methods.
The contamination from $^3_{\Lambda}{\mathrm{H}}$ decay is found to be the largest systematical source to the \dla{} measurement~\cite{Barlow:2002yb}, which yields about $13\%$ relative systematic uncertainties at $k^*\sim 10$ MeV/$c$, and reduce to less than 1\% at $k^*> 80$ MeV/$c$ for all centralities.
The uncertainty due to $\Lambda$ reconstruction purity is studied by varying topological filter parameters and $\Lambda$ selection mass window, which leads to around 6\% relative systematical uncertainties at small $k^*$ region.
The $\Lambda$ feed-down fraction $(30\pm 4)$\% is calculated by comparing thermal and transport UrQMD model calculations, and its impact on the correlation function is estimated to be less than 6\% at small $k^*$.
Other sources including $d$ identification and tracking criteria are found be less than $3\%$.
The uncertainties from different sources are then added in quadrature to obtain the total systematic uncertainties.

Figure~\ref{fig:dL_CF} shows the measured \dla{} correlation function as a function of $k^*$ for three different centralities in Au+Au collisions at $\sqrt{s_{\rm NN}}$ = 3\,GeV. 
A strong correlation is observed in the small $k^*$ region for all centralities, and the correlation strength is much larger than all the other known particle correlations measured in the same collisions (i.g. $p$-$p$, $p$-$d$, \pla{}, $p$-$\Xi$...) ~\cite{STAR:2024zvj}. 
The correlation function can be modeled as follows~\cite{Lednicky:1981su, Fabbietti:2020bfg, Lisa:2005dd}: 
\begin{equation}
    C(k^*) = \int  S(\mathbf{r^*}) |\Psi(\mathbf{r^*}, \mathbf{k^*})|^2 {\mathrm{d}^{3}}\mathbf{r^*}.
    \label{cf_theory}
\end{equation}
Here, $S(\mathbf{r^*})$ is the source function corresponding to the distribution of the relative distance of particle pairs ($r^{*}$), and  $\Psi(\mathbf{r^*}, \mathbf{k^*})$ is the pair wave function for the pairs of interest. 
Following the Lednick\'y-Lyuboshitz (LL) formalism~\cite{Lednicky:1981su} with a spherical Gaussian source under a smoothness approximation~\cite{Lisa:2005dd} convoluted with an $S$-wave function in ERE, the correlation function can be expressed with three parameters: Gaussian source ($R_G$), scattering length ($f_0$), and effective range ($d_0$). 

In this analysis, both $d$ and $\Lambda$ are treated as point-like particles and only the two-body interactions are considered. A recent calculation for \dla{} scattering using Faddeev equations for a three-body system using modern $\Lambda$-$N$ interactions shows that the LL formalism works well in the \dla{} correlation for source radii applicable to Au+Au 3 GeV collision system with small deviations seen when radii around 1.2\,fm.~\cite{Kohno:2024tkq}.

The $S$-wave \dla{} pair can have two spin states: spin-1/2 doublet (D) and spin-3/2 quartet (Q) states~\cite{Haidenbauer:2020uew, Fabbietti:2020bfg}. Quantum mechanics require the fraction of D-state and Q-state to be 1/3 and 2/3 in the wave function as follows: 
\begin{equation}
    |\Psi(\mathbf{r^*}, \mathbf{k^*})|^2=\frac{1}{3}|\Psi_{1/2}(\mathbf{r^*}, \mathbf{k^*})|^2+\frac{2}{3}|\Psi_{3/2}(\mathbf{r^*}, \mathbf{k^*})|^2.
    \label{psi_DQspin}
\end{equation}
Each state can have different strong interaction parameters ($f_0$ and $d_0$). 

\begin{table*}
    \centering
    \caption{Strong interaction parameters ($f_0$ and $d_0$) for \dla{} D and Q spin states extracted from constrained fits to real data and SMASH+LL model simulation in comparison with those predicted by various models. }
    \label{tab:f0d0}
    \begin{tabular}{c|rcl|rcl|c|c|c|c|c}
        \hline \hline
        --  & \multicolumn{3}{c|}{Data} & \multicolumn{3}{c|}{SMASH+LL} &  ~Cobis~\cite{Cobis:1996ru}~ & ~Hammer~\cite{Hammer:2001ng}~ & ~Alexander~\cite{Alexander:1968acu}~ & ~Rijken~\cite{Rijken:2010zzb}~ & ~Haidenbauer~\cite{Haidenbauer:2013oca}~ \\
        \hline
        ~$f_0 (\mathrm{D}$) (fm)~ & ~~$-26.1$ & $\pm$ & 5.6~~ & ~~ $-22.5$ & $\pm$ & 1.5~~ & ~$-16.3^{-4.0}_{+2.1}$~ & ~$-16.8^{-4.4}_{+2.4}$~ & -- & -- & -- \\
        \hline
        $d_0 (\mathrm{D}$) (fm) &  \multicolumn{3}{@{}c@{}|}{8 (at fit limit) }  & \multicolumn{3}{@{}c@{}|}{8 (at fit limit) } & 3.2 & 2.3 & -- & -- & -- \\
        \hline
        $f_0 (\mathrm{Q}$) (fm) & 18.7 & $\pm$ & 2.8 & 20.2 & $\pm$ & 1.9 & -- & -- & 7.6 & 10.8 & 17.3 \\
        \hline
        $d_0 (\mathrm{Q}$) (fm) & 6.5  & $\pm$ & 1.8 & 3.1 & $\pm$ & 0.2 & -- & -- & 3.6 & 3.8 & 3.6\\
        \hline \hline
    \end{tabular}

\end{table*}

We conduct a Bayesian inference analysis~\cite{Shen:2023awv, Mantysaari:2022ffw} and fit to experimentally measured \dla{} correlation functions within the LL formalism described above. The fit is performed simultaneously to the data in three centrality bins using three source radius parameters $R_G^{i}$ and common strong interaction parameters $f_0({\rm D})$, $d_0({\rm D})$, $f_0({\rm Q})$ and $d_0({\rm Q})$. A Monte Carlo procedure that includes the effect of experimental acceptance is incorporated for the fit to the data~\cite{Lednicky:1981su, STAR:2015kha}. Statistical and uncorrelated systematical uncertainties in the data are combined, while correlated systematic uncertainties are included in the covariance matrix used in the Bayesian analysis (see the supplemental note Sec.I).

\begin{table}
    \centering
    \caption{Gaussian source parameter (${\rm R_G}$) extracted from the LL fit to real data and SMASH+LL model simulation in different centrality bins. }
    \label{tab:source}
    \begin{tabular}{c|c|c|c|c}
    \hline \hline
        \multirow{2}{*}{~Centrality~} & \multicolumn{2}{c|}{Source Radius (fm)} & \multicolumn{2}{c}{$\langle m_T\rangle {\rm (GeV/c^{2})}$} \\ 
        & Data & Model & Data & Model\\
        \hline
        0-10\% & ~~$2.69\pm0.20$~~ & ~~$2.30\pm0.08$~~ &~~$1.68$~~ &$1.63$\\
        10-20\% & $2.44\pm0.17$ & $2.21\pm0.06$ & $1.66$ & $1.62$\\
        20-60\% & $2.08\pm0.19$ & $2.06\pm0.05$ & $1.65$ & $1.61$\\
        \hline \hline
    \end{tabular}
\end{table}

Model calculations are shown in Fig.~\ref{fig:dL_CF} in comparison to the data.
The ``Simulating Many Accelerated Strongly interacting Hadrons" (SMASH) model~\cite{SMASH:2016zqf} coupled with a coalescence procedure~\cite{Steinheimer:2012tb} is used to generate momenta and spatial coordinates of $d$ and $\Lambda$ particles~\cite{STAR:2024zvj}. Subsequently, the $d$ and $\Lambda$ momentum space correlations are modulated using the LL formalism for strong interactions with the parameters chosen to match the fitted experimental values listed in ~\Cref{tab:f0d0}. This approach allows us to investigate the effect of the collision dynamics and serves as a consistency check for the data-fitting procedure.
The correlation functions obtained from the model describe the data reasonably well, with the largest difference seen in central collisions.
We did the same Bayesian fit to the model results and the fitted parameters are shown in ~\Cref{tab:source} and ~\Cref{tab:f0d0}. The SMASH+LL model yields a $\chi^2/ndf$ of 126.8/53 when compared with data across the three centralities.
While the extracted source radii from both the data and the model show a similar trend of increasing for more central collisions, the increase is larger in real data, implying an inaccuracy in the model's simulation of collision dynamics, despite reasonably reproducing $\langle m_T \rangle$.
A similar difference is also observed in the $p$-$d$ correlation measurements~\cite{STAR:2024zvj}.
It is worthwhile to note that despite the difference in the source radii, the extracted strong interaction parameters for both spin states are consistent between data and model. This validates the factorization between the emission source and the strong interaction, and demonstrates the robustness of the obtained strong interaction parameters in this analysis approach.
The possible Coulomb effect due to the positively charged source at this energy is not included in the current LL analysis of the \dla{} correlation function. There is also lack of data on the effective Coulomb potential in various collision centralities which may be included in the future with better knowledge.

\begin{figure}[ht]
    \centering    
    \includegraphics[width=0.45\textwidth]{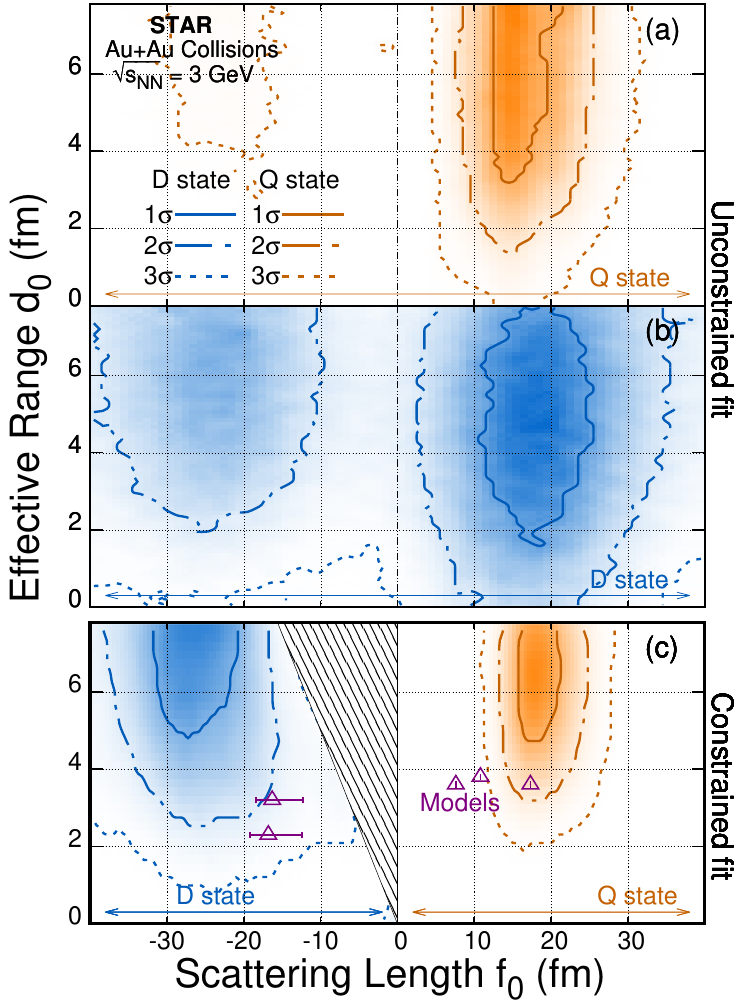}
    \caption{The extracted final state interaction parameters, scattering length ($f_0$) and effective range ($d_0$), for \dla{} interaction shown as contours of the probability distribution for both doublet (D) (blue) and quartet (Q) (orange) spin states. The unconstrained fit results are shown in panel (a) and panel (b) for Q state and D state, respectively. The constrained fit results for both states are shown in panel (c). The purple triangles show the different model calculations\cite{Cobis:1996ru, Hammer:2001ng, Alexander:1968acu, Rijken:2010zzb, Haidenbauer:2013oca}, the forbidden region is shaded. }
    \label{fig:f0d0}
\end{figure}

Panels (a) and (b) in Fig.~\ref{fig:f0d0} show the probability density distributions together with 1--3~$\sigma$ contours for D (blue) and Q (orange) spin states from the Bayesian fit, respectively. To ensure an unbiased evaluation of the fitting results without relying on prior knowledge of $^3_{\Lambda}{\mathrm{H}}$, we performed an unconstrained fit.
Both are fit within $-40$\,fm to 40\,fm for $f_0$ and 0 to 8\,fm for $d_0$, which means the two spin states are treated the same and no predefined signs are required for $f_0$ in different spin states. 
The $d_0$ parameter is limited at small values considering the strong interaction force is short range and the applicability of the EFE formulism.
A higher probability is observed at positive $f_0$ around 15 fm for both D and Q spin states. A clear hint of negative $f_0$ is also observed in the D-state, which is a necessary condition for a bound state. The minimum $\chi^2/ndf$ is found to be 85.2/53.

Conventionally, the $^3_{\Lambda}{\mathrm{H}}$ is approximated as a deuteron core surrounded by a $\Lambda$ due to its very small $\Lambda$ separation energy~\cite{Chen:2023mel}. The existence of the $^3_{\Lambda}{\mathrm{H}}$ (spin-1/2) bound state requires the D-state $f_0({\rm D})$ in \dla{} interactions to be negative~\cite{Haidenbauer:2020uew}.  
Panel (c) shows the probability density distribution with contours under a constrained fit in which the $f_0$(D) is limited to be negative ($-40<f_0({\rm D})<0$\,fm) and $f_0$(Q) is limited to be positive (0 $<f_0({\rm Q})<$ 40\,fm), while $d_0$ is still limited up to 8 fm. The results are also summarized in ~\Cref{tab:f0d0}. The hashed area is excluded from the fit with a bound state. 
The precision of the current data provides little ability to constrain the value of the $d_0$.
Sec II discusses the fit results with no constraint on $d_0$. The extracted $f_0$ from both D- and Q- states are consistent with the results shown in ~\Cref{tab:f0d0}.

The model predictions are also shown in Fig.~\ref{fig:f0d0} to compare with experimental fitting results and are listed in~\Cref{tab:f0d0}. D-state is predicted based on pionless chiral effective field theory ($\slashed{\pi}$EFT)~\cite{Hammer:2001ng} or potential models~\cite{Cobis:1996ru} with the experimentally measured $^3_{\Lambda}{\mathrm{H}}$ binding energy as input~\cite{Chen:2023mel, Haidenbauer:2020uew}. No experimental results can be referred to for a Q-state prediction, thus these predictions are calculated through $\slashed{\pi}$EFT~\cite{Alexander:1968acu}, EFT~\cite{Haidenbauer:2013oca}, or other known $Y$-$N$ potentials~\cite{Rijken:2010zzb}. This is the first time in heavy-ion collisions that we are able to extract strong interaction parameters separately for different spin states.

\begin{figure}[ht]
    \centering    
    \includegraphics[width=0.45\textwidth]{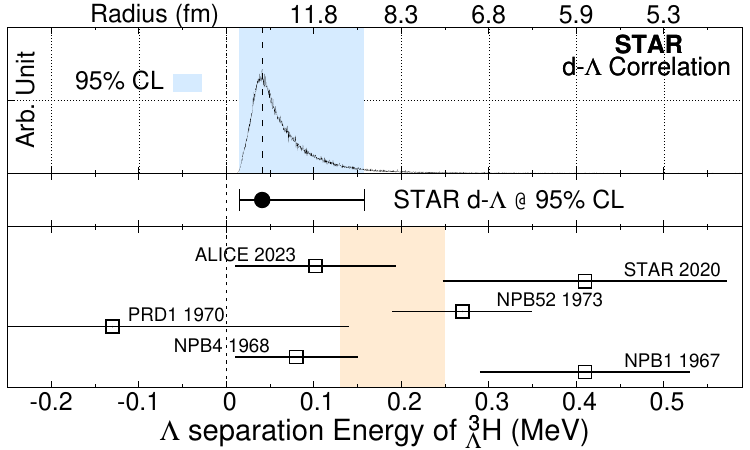}
    \caption{$\Lambda$ separation energy of $^{3}_{\Lambda}{\mathrm{H}}$ ($B_{\Lambda}$). The upper panel shows the probability distribution of the $\Lambda$ separation energy of $^{3}_{\Lambda}{\mathrm{H}}$ from \dla{} correlation. The corresponding radius is shown along the top. The middle panel shows our measured $B_{\Lambda}$ at 95\% Confidence Level (CL). Previous world-wide data~\cite{Chen:2023mel, Juric:1973zq, Keyes:1970ck, Gajewski:1967ruj, Bohm:1968qkc, STAR:2019wjm, ALICE:2022sco} are presented in the lower panel, with their weighted average of 0.19 $\pm$ 0.06 MeV illustrated by a shaded band.
    }
    \label{fig:H3LBL}
\end{figure}

The $\Lambda$ separation energy of $^3_{\Lambda}{\mathrm{H}}$, $B_{\Lambda}$, can be related to the measured D-state $f_0$ and $d_0$ in the ERE formalism by the Bethe formula $-1/f_0=\gamma-d_0\gamma^2/2$ along with $B_{\Lambda}=\gamma^2/2\mu_{\Lambda{\rm d}}$, where $\gamma$ is the binding momentum, and $\mu_{\Lambda{\rm d}}$ is the reduced mass of \dla{} pair.
Figure~\ref{fig:H3LBL} shows $B_{\Lambda}$ calculated using the $f_0$(D) and $d_0$(D) extracted in this analysis compared with previous worldwide measurements~\cite{Chen:2023mel, Juric:1973zq, Keyes:1970ck, Gajewski:1967ruj, Bohm:1968qkc, STAR:2019wjm, ALICE:2022sco}.
The upper panel shows the probability distribution from the constrained Bayesian fit, where the blue region delineates the 95\% confidence level (CL) window when including the total uncertainties.
A corresponding radius for the separation energy can be estimated by $\sqrt{\hbar^2/4\mu_{d\Lambda}B_{\Lambda}}$~\cite{Bertulani:2022vad}, and is shown along the top. We obtain $\sqrt{\langle r^2 \rangle}=18^{+12}_{-9}$\,fm at 95\% CL, comparable to that from the lifetime analysis of $^3_{\Lambda}{\mathrm{H}}$~\cite{Braun-Munzinger:2018hat,ALICE:2022sco} by the ALICE experiment at LHC and recent model predictions for a weakly bounded system~\cite{Perez-Obiol:2020qjy,Hildenbrand:2020kzu}. 
The lower panel presents prior measurements from STAR (2020)~\cite{STAR:2019wjm} and ALICE (2023)~\cite{ALICE:2022sco} using the invariant mass method from the reconstructed $^3_{\Lambda}{\mathrm{H}}$, along with four additional measurements from $K^{-}$ beam experiments~\cite{Juric:1973zq, Keyes:1970ck, Gajewski:1967ruj, Bohm:1968qkc}.
Our constrained-fit result for $B_{\Lambda}$ of $^{3}_\Lambda {\mathrm{H}}$ is $0.04_{-0.03}^{+0.12}$ MeV at 95\% CL, while our unconstrained result (discussed in Sec.II) is $0.03_{-0.02}^{+0.25}$\,MeV at 95\% CL. Both of these results are consistent with the previous weighted world average of $0.19\pm0.06$\,MeV, shown as a shaded band in Fig.~\ref{fig:H3LBL}.

To summarize, the first observation of \dla{} correlation in Au+Au collisions at $\sqrt{s_{\rm NN}}$ = 3\,GeV is reported in this Letter. 
With the Lednick\'y-Lyuboshitz approach, the spin-dependent strong interaction parameters ($f_0$ and $d_0$) are separated from the emission source.
A growth in the measured source size $R_G$ is seen with increasing centrality in the real data, while calculations from the hadronic transport model SMASH, coupled with a coalescence procedure for deuteron formation, underestimate this growth.
Although the unconstrained fit results favor final state interaction parameter $f_0 > 0$,
a clear hint of $f_0<0$ for the D-state is seen, consistent with the formation of $^{3}_\Lambda \mathrm{H}$. 
A constrained fit yields 
$f_0({\rm D})=-26.1 \pm 5.6$ fm and $f_0({\rm Q})=18.7\pm2.8$ fm, for the D and Q spin states respectively. 
From the scattering parameters of the D-state, the $\Lambda$ separation energy of $^{3}_\Lambda \mathrm{H}$ is found to be $0.04_{-0.03}^{+0.12}$ MeV at 95\% confidence level. 
Future experimental measurements on nucleon/nuclei-hyperon correlations with improved statistical precisions and more theoretical
developments including multi-body interactions are needed to further deepen the knowledge of hyperon-nucleon interactions and 
 their roles within the QCD EoS in the region of high density.

We thank the RHIC Operations Group and SDCC at BNL, the NERSC Center at LBNL, and the Open Science Grid consortium for providing resources and support.  This work was supported in part by the Office of Nuclear Physics within the U.S. DOE Office of Science, the U.S. National Science Foundation, National Natural Science Foundation of China, Chinese Academy of Science, the Ministry of Science and Technology of China and the Chinese Ministry of Education, NSTC Taipei, the National Research Foundation of Korea, Czech Science Foundation and Ministry of Education, Youth and Sports of the Czech Republic, Hungarian National Research, Development and Innovation Office, New National Excellency Programme of the Hungarian Ministry of Human Capacities, Department of Atomic Energy and Department of Science and Technology of the Government of India, the National Science Centre and WUT ID-UB of Poland, the Ministry of Science, Education and Sports of the Republic of Croatia, German Bundesministerium f\"ur Bildung, Wissenschaft, Forschung and Technologie (BMBF), Helmholtz Association, Ministry of Education, Culture, Sports, Science, and Technology (MEXT), Japan Society for the Promotion of Science (JSPS) and Agencia Nacional de Investigaci\'on y Desarrollo (ANID) of Chile. We thank Drs. Chun Shen and Wenbin Zhao for the fruitful discussion.

\bibliographystyle{apsrev4-2}
\bibliography{apssamp}

\end{document}


\title{Supplemental Note: First observation of deuteron-{$\Lambda$} correlations at RHIC }
\author{(STAR Collaboration)}
\date{\today}
\maketitle

\section{Bayesian Inference Analysis}
\label{sec-error-matrix}
Bayesian analysis provides a probabilistic framework for parameter estimation by combining prior knowledge (in this case, the LL model~\cite{Lednicky:1981su}) with observational data. It is particularly effective in handling multiparameter fitting problems. In this work, we trained emulators with our LL model calculations using 8000 design points in the model parameter space, sampled via the Maximum Projection Latin Hypercube Design algorithm~\cite{10.1093/biomet/asv002}. With these trained emulators, we can efficiently obtain the posterior distribution $P(\theta | y_{\mathrm{exp}})$ of the model parameters by sampling a uniform prior $P(\theta)$ using the Markov Chain Monte Carlo (MCMC) method. The Bayesian inference framework follows references~\cite{Shen:2023awv, Mantysaari:2022ffw}, and we have implemented covariance error fitting within the original framework to better account for experimental data uncertainties. 

\begin{figure}[ht]
    \centering
    \includegraphics[width=\linewidth]{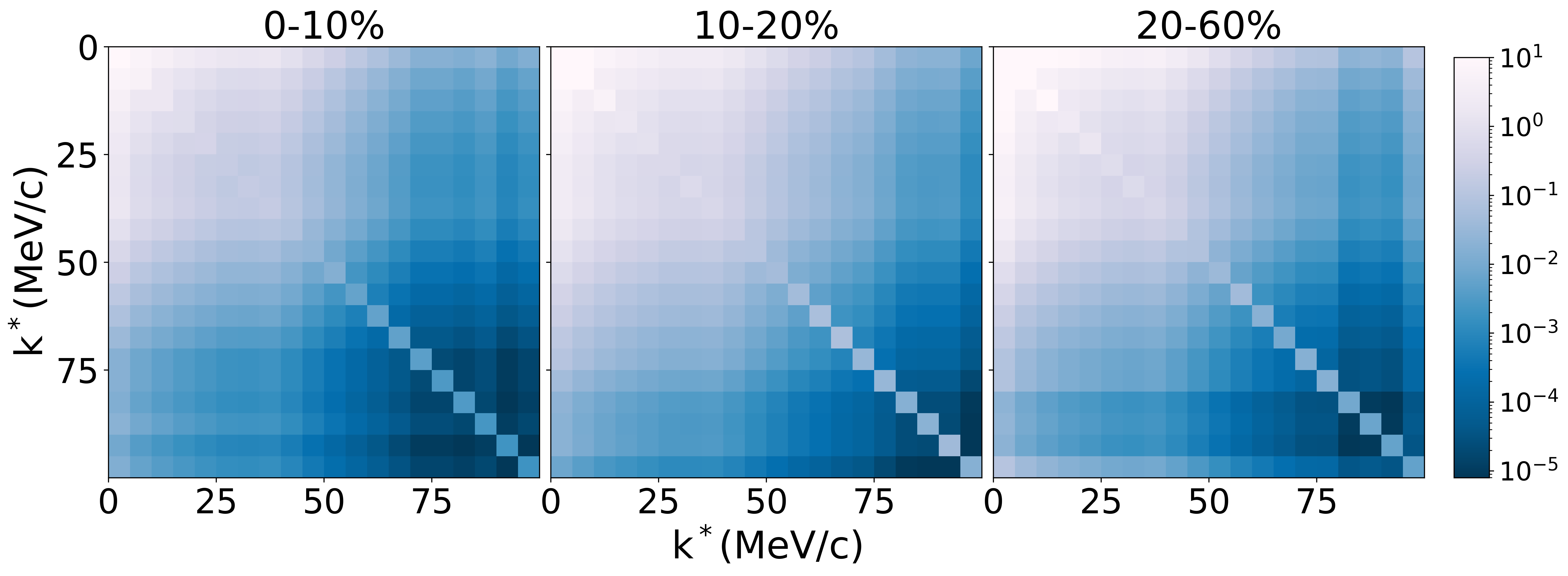}
    \caption{The covariance matrix of correlated and uncorrelated uncertainties for \dla{} correlation functions, shown for 0-10\% (left), 10-20\% (middle), and 20-60\% (right) centralities. The diagonal terms represent the total uncertainty, combining correlated and uncorrelated uncertainties, while the off-diagonal terms represent the correlated uncertainties between data points.}
    \label{fig:err-matrix}
\end{figure}

The uncertainties discussed in the main text are calculated as point-by-point correlated uncertainties ($\sigma_{i}^{\rm corr}$) and uncorrelated uncertainties ($\sigma_{i}^{\rm uncorr}$). The correlated uncertainties include contributions from $^3_{\Lambda}{\mathrm{H}}$ three-body decay contamination and the feed-down to $\Lambda$ from decays of other strange baryons.
The uncorrelated systematic uncertainties are combined with the statistical uncertainties in quadrature as $\sigma_{i}^{\rm uncorr}$.
Figure~\ref{fig:err-matrix} shows the uncertainty covariance matrix of the \dla{} correlation function data. For each row ($i$, start from top) and column ($j$, start from left), the diagonal term ($i=j$) is sum of the uncorrelated and correlated uncertainty ($(\sigma_{i}^{\rm corr})^2+(\sigma_{i}^{\rm uncorr})^2$), the non-diagonal term ($i\ne j$) is the correlation between two data points in the same centrality ($\sigma_{i}^{\rm corr}\sigma_{j}^{\rm corr}$). 

\begin{figure}[ht]
    \centering
    \includegraphics[width=\linewidth]{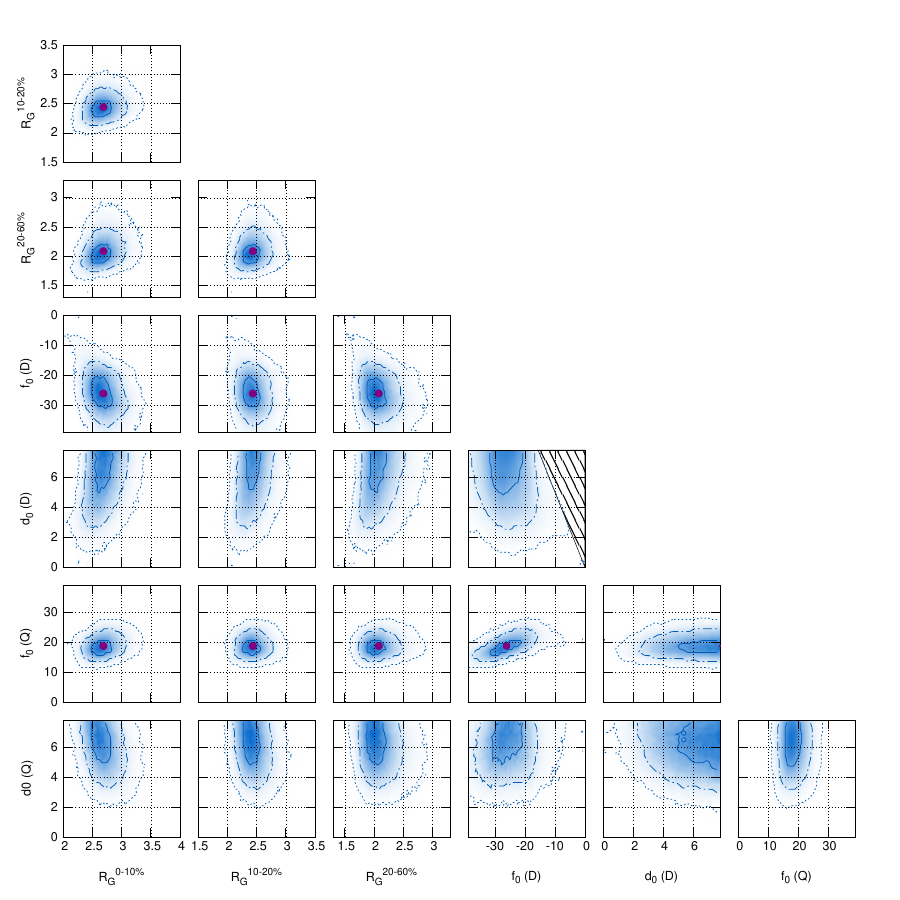}
    \caption{Posterior distributions of the extracted source size parameter (${\rm R_G}$) and final-state interaction parameters ($f_0$ and $d_0$ for both doublet (D) and quartet (Q) spin states) for the \dla{} interaction, shown as probability contours obtained from the Bayesian inference analysis. Each panel illustrates correlations between two parameters, with contour lines indicating the $1\sigma$, $2\sigma$, and $3\sigma$ confidence intervals. The purple points represent the best-fit values, and the shaded areas denote physically forbidden regions.}
    \label{fig:corner_plot}
\end{figure}

The posterior distributions for constrained fit are shown in Fig.~\ref{fig:corner_plot} for correlations between each two pair of parameters (${\rm R_G^{0-10\%}}$, ${\rm R_G^{10-20\%}}$, ${\rm R_G^{20-60\%}}$, $f_0$ (D), $d_0$ (D), $f_0$ (Q), and $d_0$ (Q),), and the correlations for $f_0$ and $d_0$ are shown in the main text for discussion. A closure test with the SMASH+LL model is applied to verify the whole framework further.

\section{Fit with unconstrained effective range parameter ($d_0$)}
\label{sec-d0}
\begin{figure}[ht]
    \centering
    \includegraphics[width=0.9\linewidth]{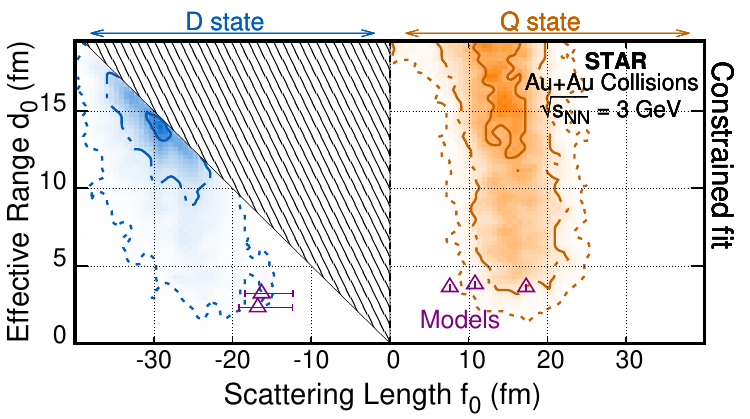}
    \caption{The extracted final state interaction parameters, scattering length ($f_0$) and effective range ($d_0$), for \dla{} interaction shown as contours of the probability distribution for both doublet (D) (blue) and quartet (Q) (orange) spin states when constraining only $f_0$ (no limit on $d_0$). The purple triangles show the different model calculations noted in Table~\ref{tab:f0d0}~\cite{Cobis:1996ru, Hammer:2001ng, Alexander:1968acu, Rijken:2010zzb, Haidenbauer:2013oca}.}
    \label{fig:f0d0-20}
\end{figure}

While the primary discussion of this Letter has focused on fits that limit $d_0$ due to the current statistics' poor ability to constrain its value in the fits, we include here and in Fig.~\ref{fig:f0d0-20} for completeness our findings when only $f_0$ is constrained.
The scattering length and effective range are found to be $f_0({\rm D}) = -29.2\pm4.3$ fm, $d_0({\rm D}) = 13.3\pm2.2$ fm, $f_0({\rm Q}) = 14.9\pm3.4$ fm. $d_0({\rm Q})$ is unconverged even up to 20 fm. The corresponding $\Lambda$ separation energy of $^{3}_\Lambda \mathrm{H}$ is determined to be $0.03_{-0.02}^{+0.25}$\,MeV at 95\% CL.

\bibliographystyle{apsrev4-2}
\bibliography{apssamp}